\documentclass[aps,preprint,showpacs,11pt,floatfix]{revtex4}
\usepackage{graphics}

\begin{document}
\title{Slowly rotating neutron stars and hadronic stars in chiral SU(3) quark mean field model}

\author{Shaoyu Yin$^1$\footnote{051019008@fudan.edu.cn}, Jiadong Zang$^1$,
Ru-Keng Su$^{1,2,3}$\footnote{rksu@fudan.ac.cn}}\affiliation{\small
1. Department of Physics, Fudan University, Shanghai 200433, People's Republic of China\\
\small 2. China Center of Advanced Science and Technology (World
Laboratory), P.B.Box 8730, Beijing 100080, People's Republic of
China\\
\small 3. Center of Theoretical Nuclear Physics, National Laboratory
of Heavy Ion Collisions, Lanzhou 730000, People's Republic of China}

\begin{abstract}
The equations of state for neutron matter, strange and non-strange
hadronic matter in a chiral SU(3) quark mean field model are applied
in the study of slowly rotating neutron stars and hadronic stars.
The radius, mass, moment of inertia, and other physical quantities
are carefully examined. The effect of nucleon crust for the strange
hadronic star is exhibited. Our results show the rotation can
increase the maximum mass of compact stars significantly. For big
enough mass of pulsar which can not be explained as strange hadronic
star, the theoretical approaches to increase the maximum mass are
addressed.
\end{abstract}


\maketitle

\section{introduction}

Stars are not only the focus of astronomers but also perfect
laboratories for theoretical physicists, since extreme stars can
supply conditions hard to be realized in earth-based laboratories.
The neutron star was one of the most active actors on the stage of
astronomy and physics since 1960s, when the first pulsar was
observed \cite{Hewish:1968}, and is still attracting more
concentration. Many different aspects of neutron stars have been
examined, such as the moment of inertia \cite{Ravenhall:1994,
Kalogera:1999}, the early stage of proto-neutron star
\cite{Prakash:1997}, the thermodynamic structure \cite{Gupta:2007},
\textit{etc.} However, because of the lack of the knowledge on the
details of the matter at high density, the true equation of state
(EoS) in such extreme object is still unclear. Great effort to
determine the EoS of compact stars at extremely high density have
been made by many authors, either by the theoretical calculation
\cite{Baym:1971,Glendenning:1985,Wu:2009} or by observational
constraint \cite{Lattimer:2001,Klahn:2006,Lattimer:2007}. At first,
considering the extremely heavy mass and strong pressure, people
regarded the pulsars as rotating neutron stars for a long time.
However, after the pioneer work of Witten \cite{Witten:1984} that
the strange quark matter is comparatively more stable than the
normal nucleus, the strange quark star \cite{Haensel:1986,
Alcock:1986,Xu:2003,Shen:2005} have attracted much attention (for a
comprehensive review, see Ref.\cite{Weber:2005}). Up to now, there
is still no evidence strong enough to confirm the existence of
strange quark star. The question is still open and need more careful
study.

In fact, the neutron star and the strange quark star are just very
simple categories, and the real components in the star should be
much more complicated. It is generally believed that the $\beta$
equilibrium can be achieved in the process of neutron-star formation
and the hyperons will exist in neutron star. The consideration of
hyperons in the EoS of compact star was back to as early as 1960
\cite{Ambartsumyan:1960} and got serious discussion in many later
references \cite{Glendenning:1982,Glendenning:1985,Schaffner:1996,
Baldo:dual}. The existence of hyperons can affect the EoS and
consequently the mass-radius ($M$-$R$) relation remarkably.

In this paper, we will concentrate on a chiral SU(3) quark mean
field model \cite{Wang:2001,Wang:2002}, which has been applied
successfully to the study of nuclear matter, strange hadronic
matter, nuclei and hypernuclei. In this model, quarks are confined
within baryons and interact with mesons. The self-interaction of
mesons is based on the SU(3) chiral symmetry. This model has been
employed by Ref.\cite{Wang:2005} to discuss the properties of static
(non-)strange hadronic stars and neutron stars in $\beta$
equilibrium. However, the resulting maximum mass of the strange
hadronic star is only $1.45M_{sun}$, which is incapable to explain
the existence of some massive stars, such as the recently confirmed
data of PSR J1903+0327 with $M=1.67\pm0.01M_{sun}$
\cite{Freire:2009}. However, we notice that the study in
Ref.\cite{Wang:2005} limited to the static case. But in fact,
rotation is a very general property of almost all stellar bodies,
for example, PSR J1903+0327 has a very short rotating period
$2.15$ms. This motivates us to extend the discussions of
Ref.\cite{Wang:2005} to rotating cases and examine the rotation
effect on the maximum mass. In this paper we will utilize Hartle's
method \cite{Hartle:dual} to study the slowly rotating neutron star
and hadronic star with the EoS given by Ref.\cite{Wang:2005}. We
will also compare the $M$-$R$ relation and other properties with
those of the strange quark star obtained by the quark mass density-
and temperature-dependent (QMDTD) model in Ref.\cite{Shen:2005}.

The organization of this paper is as follows. In the next section,
we will show the frame of Hartle's method applied in our
calculation. In Sec. III, we will take the strange hadronic EoS of
the chiral SU(3) quark mean field model to study the slowly rotating
strange hadronic star. The effect of rotation on the maximum mass
and moment of inertia will be discussed in detail. In Sec. IV, we
will briefly show the numerical results of the non-strange hadronic
and neutron star. Together with the results of the strange hadronic
star in Sec. III and the strange quark star in Ref.\cite{Shen:2005},
all these results for different stars will be compared. The last
section is for summary and discussion.

\section{Theoretical formalism for slowly rotating star}

First we will present the formalism of Hartle's method on the slowly
rotating star. This method treats the slow rotation as a small
perturbation to the non-rotating structure. In static frame, the
metric of a non-rotating, spherically symmetric star is given by
\begin{equation}
ds^2=-e^{\nu(r)}dt^2+e^{\lambda(r)}dr^2+r^2(d\theta^2+\sin^2\theta
d\varphi^2).
\end{equation}
where $e^{\lambda(r)}=\left(1-\frac{2m(r)}{r}\right)^{-1}$,
$m(r)=\int_0^r4\pi r'^2\rho(r')dr'$. In hydrostatic equilibrium, the
coefficients $\lambda$ and $\nu$ are governed by the TOV equations:
\begin{equation}\label{TOV1}
\frac{d\nu(r)}{dr}=-\frac{2}{\rho(r)+P(r)}\frac{dP(r)}{dr},
\end{equation}
\begin{equation}\label{TOV2}
\frac{dP(r)}{dr}=-\frac{[P(r)+\rho(r)][m(r)+4\pi
r^3P(r)]}{r^2\left[1-\frac{2m(r)}{r}\right]},
\end{equation}
where $P(r)$ and $\rho(r)$ are the pressure and the energy density
of the matter at radius $r$, respectively. The surface of the star
$r=R$ is defined by $P(R)=0$ and the stellar mass is $M=m(R)$.
Outside the star, the metric is simply of the Schwarzschild form,
$e^{\nu(r)}=1-\frac{2M}{r}$, so the inner and outer solution must
connect at the surface of the star. These equations can be solved
numerically by integrating outward from $r=0$ with a given center
pressure $P(0)=P_c$ till the surface where $P=0$ is reached, then
the main structure of a non-rotating star is obtained.

Now we move ahead to the slowly rotating star with angular velocity
$\Omega(r,\theta)$. The metric is not static and isotropic any
longer, and the crossing term $g_{t\theta}$ in the metric emerges.
We will now apply Hartle's method. Since this method had been shown
in detail in Ref.\cite{Hartle:dual}, hereafter we only write down
the essential steps which are necessary in our numerical
calculation. Following Hartle's expression, noticing $\Omega$ is
very small, the metric can be expanded up to the second order of
$\Omega$:
\begin{eqnarray}\label{metric}
ds^2&=&-e^{\nu(r)}[1+2h(r,\theta)]dt^2+e^{\lambda(r)}
\left[1+\frac{2\tilde{m}(r,\theta)}{r-2m}\right]dr^2\nonumber\\
&&+r^2[1+2k(r,\theta)][d\theta^2+\sin^2\theta(d\varphi-\omega(r,\theta)
dt)^2]+O(\Omega^3),
\end{eqnarray}
where $h$, $\tilde m$ and $k$ are corrections to the non-rotating
metric of order $\Omega^2$, and $\omega(r,\theta)$ is linear order
of $\Omega$ in the $g_{t\theta}$ component, which physically stands
for the angular velocity of the local inertial frame. Define
$\varpi(r,\theta)=\Omega(r,\theta)-\omega(r,\theta)$, the angular
dependence can be studied by the vector spherical harmonics
expansion \cite{Regge:1957}. However, for the uniform rotating case
$\Omega(r,\theta)=\Omega$, one can find that $\varpi$ is a function
of $r$ alone as required by the boundary condition, so the equation
obeyed by $\varpi$ is then simply \cite{Hartle:dual}
\begin{equation}
\frac{1}{r^4}\frac{d}{dr}\left[r^4j(r)\frac{d\varpi(r)}{dr}\right]+
\frac{4}{r}\frac{dj(r)}{dr}\varpi(r)=0,
\end{equation}
where $j(r)=e^{-\frac{\nu(r)+\lambda(r)}{2}}$. The boundary
condition at $r=0$ is $\frac{d\varpi}{dr}=0$, while outside the
star, $j(r)\equiv1$ and the solution is
\begin{equation}\label{relat}
\varpi(r)=\Omega-\frac{2J}{r^3},
\end{equation}
where $J$ is the total angular momentum of the star. The moment of
inertia is given by
\begin{equation}
I=\frac{J}{\Omega}.
\end{equation}
Besides the metric, the pressure is also affected by the rotation,
which is denoted by an dimensionless parameter $p^*$:
\begin{equation}
p^*=\Delta P/(P+\rho).
\end{equation}
The metric perturbation terms as well as $p^*$ can be expanded in
spherical harmonics:
\begin{eqnarray}
&&h(r,\theta)=h_0(r)+h_2P_2(\theta)+\cdots,\\
&&\tilde{m}(r,\theta)=\tilde{m}_0(r)+\tilde{m}_2P_2(\theta)+\cdots,\\
&&k(r,\theta)=k_0(r)+k_2P_2(\theta)+\cdots,\\
&&p^*(r,\theta)=p^*_0(r)+p^*_2(r)P_2(\theta)+\cdots,
\end{eqnarray}
where the odd terms automatically vanish because of the symmetry.
The field equation together with the TOV equations yields the
equations for the $l=0$ correction terms:
\begin{eqnarray}
&&\frac{d\tilde m_0}{dr}=4\pi r^2p_0^*(\rho+P)\frac{d\rho}{dP}
+\frac{1}{12}j^2r^4\left(\frac{d\varpi}{dr}\right)^2
-\frac{1}{3}r^3\frac{dj^2}{dr}\varpi^2,\\
&&\frac{dp_0^*}{dr}=\frac{1}{12}\frac{r^4j^2}{r-2m}\left(
\frac{d\varpi}{dr}\right)^2+\frac{1}{3}\frac{d}{dr}\left(\frac{r^3j^2
\varpi^2}{r-2m}\right)-\frac{4\pi(\rho+P)r^2}{r-2m}p_0^*-\frac{\tilde
m_0(1+8\pi r^2P)}{(r-2m)^2},
\end{eqnarray}
where the boundary conditions are $p_0^*(0)=\tilde{m}(0)=0$. Given a
center rotating velocity $\varpi(0)=\varpi_c$, the above formalism
is enough to determine the changes on the stellar mass $\delta M$
and the mean radius $\delta R$,
\begin{eqnarray}
&&\delta M=\tilde m_0(R)+\frac{J^2}{R^3},\\
&&\delta R=-p^*_0(R)\rho(R)\bigg/\frac{dP(R)}{dr}.
\end{eqnarray}
Higher order terms of the metric corrections are related to the
configuration distortion away from an exact spheroid, which is
neglected in our study. Under the linear approximation, we can use
the relation $\varpi_c'=\varpi_c\frac{\Omega'}{\Omega}$ to obtain
the desired value of $\varpi_c'$ at the center from any initial
value $\varpi_c$, where $\Omega'$ and $\Omega$ are the corresponding
angular velocities.

Finally we have to mention the so-called Kepler limit. With the
increasing frequency, the tangent velocity of the matter on the
stellar surface is also increasing. It is the gravitational force
which keeps the matter from escaping. However, when the angular
velocity is beyond a critical value, the gravitational attraction
becomes insufficient to balance the centrifugal force, and the star
becomes dynamically unstable. This onset is the Kepler limit given
by the empirical formula
\begin{equation}\label{Kepler}
\Omega_K(M)\approx C(M/M_{sun})^{1/2}(R/10\textrm{km})^{-3/2},
\end{equation}
where $R$ corresponds to the radius of a non-rotating star with mass
$M$, and $C=1.04$kHz is a constant independent of the EoS
\cite{Lattimer:2004}. When the rotating frequency approaches the
Kepler limit, the stellar equator coincides with the innermost
rotating orbit and the surface mass will shed away. Therefore we
should always restrict our reasonable results within the Kepler
limit.

\section{Numerical results of strange hadronic star}

With Hartle's formalism, we can calculate the physical properties
such as the mass, radius and angular momentum of a slowly rotating
star with any given EoS. In this section we apply the EoS of strange
hadronic matter from the chiral SU(3) quark mean field model
\cite{Wang:2005} to examine the properties of slowly rotating
strange hadronic star.

Before heading on to the numerical results, we should say a few
words about the EoS at lower density. It is commonly believed that
the neutron stars or quark stars are enveloped by a nucleon crust at
the surface, where the density and pressure are much lower than in
the center. The nucleon crust is supposed to be responsible for the
glitch of the rotating frequency \cite{Anderson:1975,
Glendenning:1992a,Zdunik:2001}, and it can also affect the stellar
radius significantly. To examine the property of the crust
explicitly in the rotating hadronic star, we will compare the
results of the strange hadronic star with and without the crust. The
crust EoS from different models are generally similar to each other.
In order to compare the rotating results with those of the
non-rotating case given by Ref.\cite{Wang:2005}, for
$\rho_B<0.1$fm$^{-3}$, we still use the same nuclear EoS as that of
Ref.\cite{Wang:2005}, namely, the EoS first given by
Ref.\cite{Negele:1974}, though many other good EoS for nuclear
matter existed \cite{Baym:1971,Ruster:2006}. At
$\rho_B>0.1$fm$^{-3}$, the EoS of strange hadronic matter shown in
Fig.4 of Ref.\cite{Wang:2005} will be used.

\begin{figure}
\resizebox{0.7\linewidth}{!}{\includegraphics*{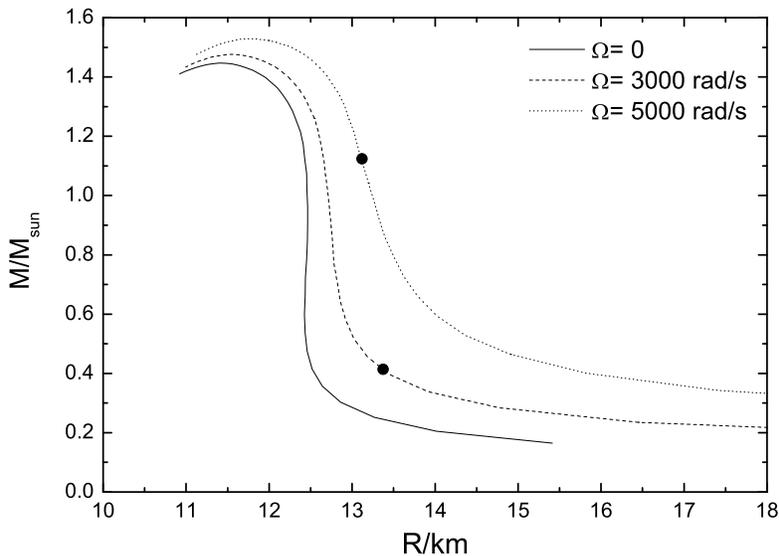}}
\caption{The $M$-$R$ relation of strange hadronic star with crust at
different angular velocities, where the round dots denote the
minimum stable mass as required by the Kepler
instability.}\label{mrcr}
\end{figure}

In Fig.\ref{mrcr} we show the $M$-$R$ curves of crusted strange
hadronic stars at different angular velocities. The curve turning
over the peak value of $M$ is kept with only a very short segment,
since it is well known that the star becomes unstable at this
segment. As can be clearly seen, when the rotation is imposed, the
mass and radius are increased significantly. The faster the star
rotates, the heavier and larger it can be. The round dot on each
curve denotes the minimum stable mass at the corresponding angular
velocity, which is obtained by the criterion of the Kepler limit
following Eq.(\ref{Kepler}). Hereafter we will always use such round
dots to denote the minimum stable mass on each curve, the segment
with mass lower than the dot is unstable because of the violation on
the Kepler limit. We also see from Fig.\ref{mrcr} that the increase
in the radius of the maximum mass at $\Omega=5000$rad/s from the
non-rotating case is about $3.55\%$ only, so the shape deformation
is indeed small. This confirms that our application of Hartle's
method is self-consistent. In fact, as was demonstrated by Weber and
Glendenning \cite{Weber:1992}, Hartle's method is applicable to
stars with the rotation period up to $0.5$ms, or in our language
$\Omega=12566$rad/s, which is much larger than the fastest rotation
speed observed yet.

\begin{figure}
\resizebox{0.7\linewidth}{!}{\includegraphics*{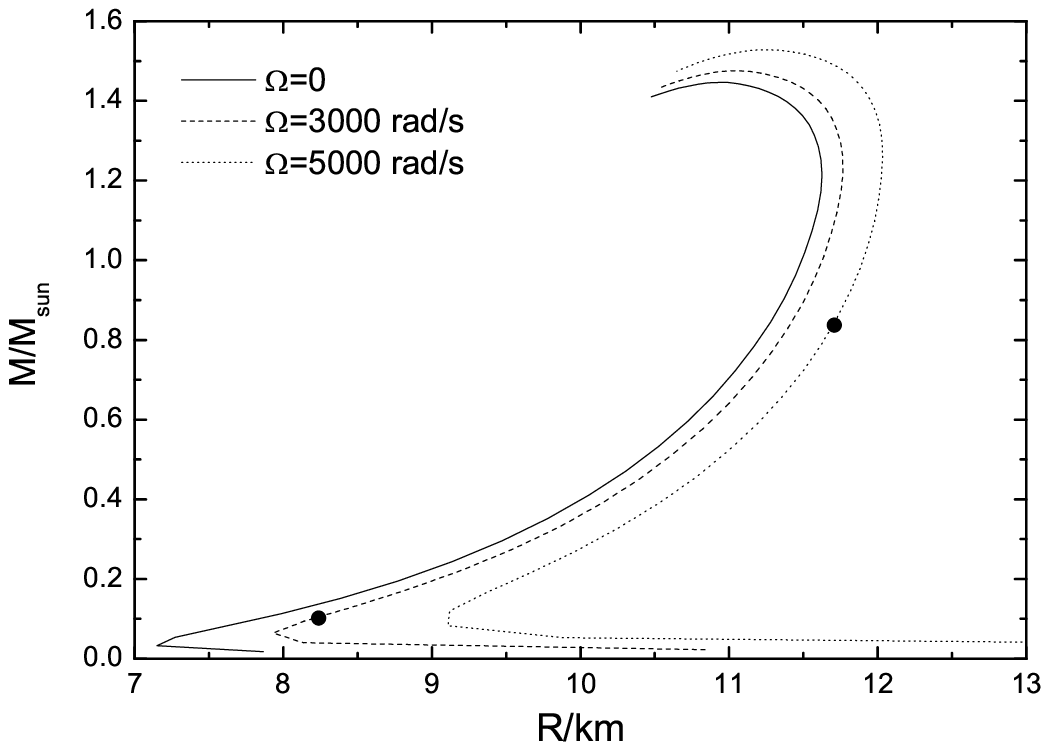}} \caption{The
same as Fig.\ref{mrcr}, but the crust is not considered.}\label{mr}
\end{figure}

To illustrate the effect of the crust clearly, we also plot the
corresponding results without crust in Fig.\ref{mr}, where the EoS
of the chiral SU(3) model is still used when $\rho_B<0.1$fm$^{-3}$.
It is clearly shown that the shape of the $M$-$R$ curve is
dramatically changed. For a large range of $M$ about
$0.1\sim1.2M_{sun}$, the $M$-$R$ curves have different direction
compared to the case with the crust in Fig.\ref{mrcr}. When the
crust of lower density is considered, the radius of the star can be
quite large when $M$ is small, to which the outer crust contributes
the most.

\begin{figure}
\resizebox{0.7\linewidth}{!}{\includegraphics*{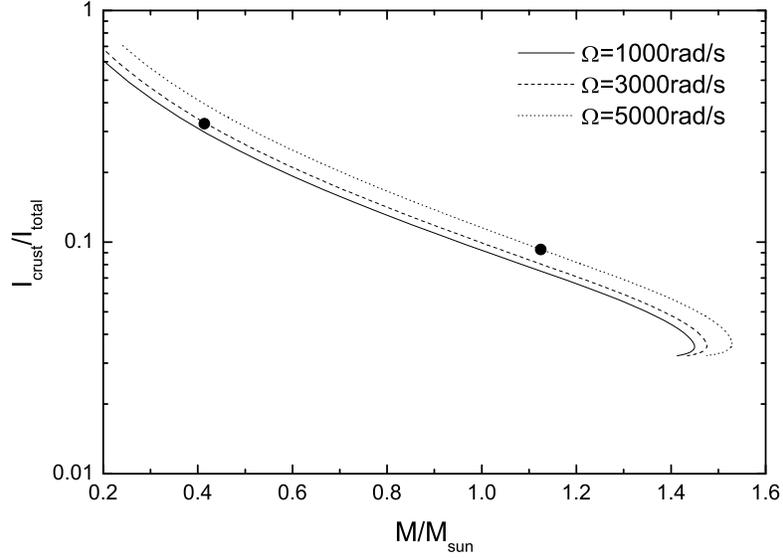}}
\caption{The crust portion of the moment of inertia as a function of
stellar mass at different rotation speeds.}\label{Icrm}
\end{figure}

Another important property about the crust is its moment of inertia,
which is commonly used to explain the glitch of the rotating
frequency. The glitch is a sudden change of the angular frequency,
which can be considered as the transfer of the angular momentum
between the crust and inner part of the star. The recent observation
requires that the crust must contain at least $1.4\%$ of the total
moment of inertia \cite{Lattimer:2001}. Now we calculate the ratio
of the moment of inertia of the crust to that of the whole strange
hadronic star. Since the moment of inertia can be calculated by
\begin{equation}
I=\frac{1}{\Omega}\int T_3^0\sqrt{-g}dV,
\end{equation}
in our first order spherical approximation, it can be expressed as
\begin{equation}
I\approx\frac{3\pi}{8}\int_0^R\frac{e^{-\nu(r)/2}r^4[(\rho(r)+p(r)]}
{\sqrt{1-\frac{2m(r)}{r}}}\frac{[\Omega-\omega(r)]}{\Omega}dr.
\end{equation}
The crust portion can be easily calculated by substituting the lower
limit of the integration from $0$ to $R_{crust}$, which is
determined by $\rho_B(R_{crust})=0.1$fm$^{-3}$. In Fig.\ref{Icrm},
we show the crust portion of the moment of inertia in the total star
at different rotation speeds. We see that the crust contributes
about $3.5\%$ to the total moment of inertia at maximum mass, and
this portion is larger at smaller mass. So the requirement of
Ref.\cite{Lattimer:2001} is well satisfied.

\begin{figure}
\resizebox{0.7\linewidth}{!}{\includegraphics*{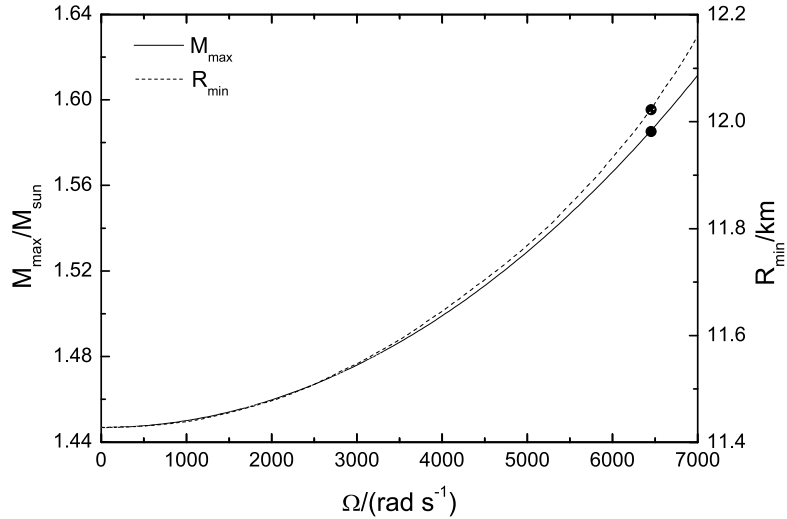}}
\caption{The maximum mass and the minimum stable radius of strange
hadronic star {\it vs.} the angular velocity.} \label{MRW}
\end{figure}

The most important thing we are interested in is the influence of
the angular velocity on the maximum mass, which is shown in
Fig.\ref{MRW}, where the curve of minimum stable radii {\it vs.} the
angular velocity is also plotted. The minimum stable radius
corresponding to the maximum mass can be found directly from the
$M$-$R$ curve. We find that the minimum stable radius increases with
$\Omega$. The dots denote the Kepler limit $\Omega=6454$rad/s
obtained from the formula in Eq.(\ref{Kepler}). It is the maximum
rotation speed corresponding to the maximum mass of the non-rotating
$M$-$R$ curve. As can be seen from the $M$-$R$ curves in
Fig.\ref{mrcr}, since the $M$-$R$ curve of rotating star is higher
than the non-rotating case, the maximum mass on the $M$-$R$ curve
with such maximum rotation speed dose not lie on the boundary of
Kepler limit, so this maximum mass can exist safely and lies far
from the onset of dynamical instability.

\begin{figure}
\resizebox{0.7\linewidth}{!}{\includegraphics*{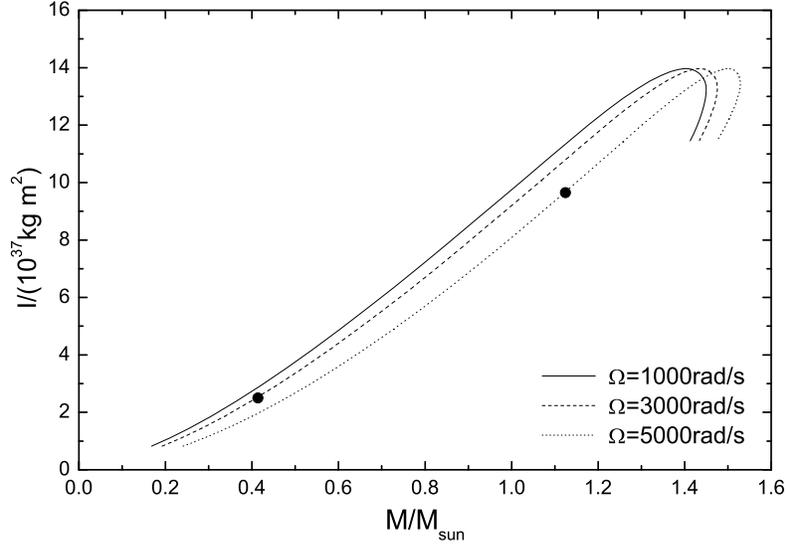}} \caption{The
moment of inertia $I$ {\it vs.} stellar mass $M$ of strange hadronic
star at different angular velocities.}\label{Im}
\end{figure}

Besides the $M$-$R$ curve, the $I$-$M$ curve is also very important
because it contains information about the inner property of compact
stars. We show the $I$-$M$ relation at different angular velocities
in Fig.\ref{Im}.

\section{Comparison between different stars}

\begin{figure}
\resizebox{0.7\linewidth}{!}{\includegraphics*{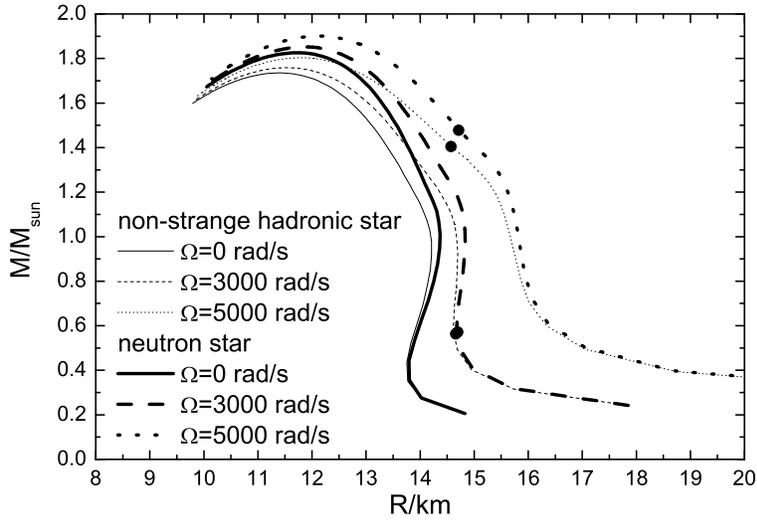}}
\caption{The $M$-$R$ curves of the non-strange hadronic star and the
neutron star at different rotation speeds.}\label{mrtwo}
\end{figure}

\begin{figure}
\resizebox{0.7\linewidth}{!}{\includegraphics*{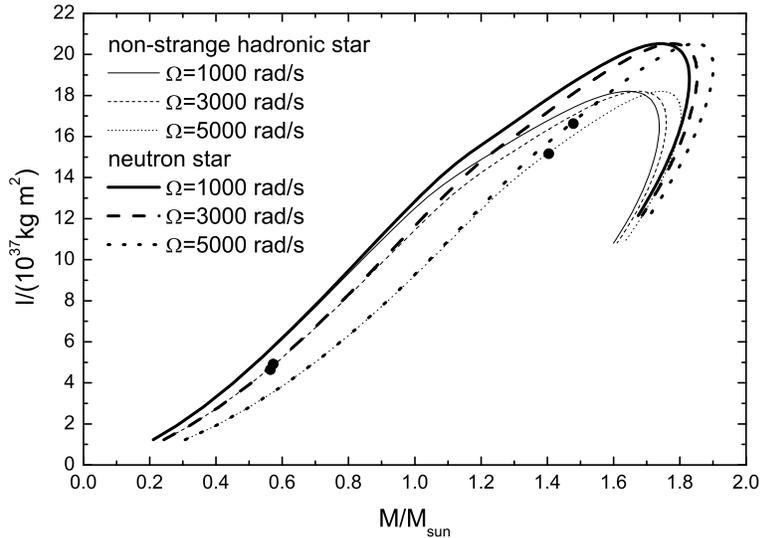}}
\caption{The $I$-$M$ curves of the non-strange hadronic star and the
neutron star at different rotation speeds.}\label{Imtwo}
\end{figure}

In previous section we have shown the numerical results of the
slowly rotating strange hadronic star in detail. Similarly, we can
discuss the properties of the rotating non-strange hadronic star and
neutron star. Here ``non-strange" means that no hyperon is
considered in the EoS, while ``neutron" means only the neutron is
considered. These EoS are also given in Fig.4 of
Ref.\cite{Wang:2005}, while the nucleon crust will always be
considered from now on. The $M$-$R$ curves for the non-strange
hadronic star and the neutron star are shown in Fig.\ref{mrtwo}, and
the $I$-$M$ curves in Fig.\ref{Imtwo}, respectively. All the Kepler
limit points are dotted as in Sec.III. We see from these figures
that the qualitative behavior is the same as the strange hadronic
star. The masses and moments of inertia in these two cases are
larger than the strange hadronic star, and the neutron star has the
largest $M$ and $I$ among these three types. Another remarkable
property shown in Figs.\ref{mrtwo} and \ref{Imtwo} is that the
curves of the same type are closer at bigger mass, while at smaller
mass the curves of the same angular velocity are closer. This means
that the rotational effect dominates in the lighter star with
smaller central pressure.


With all these results, now we can make some comparisons between
different stars. They are the strange hadronic star, non-strange
hadronic star and the neutron star obtained by the chiral SU(3)
quark mean field model, as well as the strange quark star from the
QMDTD model, whose properties are carefully studied in
Ref.\cite{Shen:2005}.

A very important consistency check is that these stars have similar
behavior according to the effect of the rotation. This is quite
reasonable since the dynamical properties of a compact star are
qualitatively similar. The centrifugal force from the rotation
counteracting the gravitation can help the star bearing more mass
and extending to larger radius. The details of these properties have
been discussed clearly in Sec. III.

One significant distinctness of the strange quark star is that its
$M$-$R$ curve has different direction from the other three types at
small $M$, as was shown in Fig.2 of Ref.\cite{Shen:2005}. The reason
is that the crust is not considered in Ref.\cite{Shen:2005}, so
similar behavior is observed in our Fig.\ref{mr}. When the star is
very small, the crust thickness becomes important, which will change
the shape of the $M$-$R$ curve significantly. Generally, the nucleon
crust can increase the radius by about $1$km but has very little
influence on the maximum mass. One difference between Fig.2 of
Ref.\cite{Shen:2005} and our Fig.\ref{mr} is that the radius
approaches to zero with decreasing $M$ for the bare strange quark
star, while it finally goes very large in our results of bare
strange hadronic star. This difference can be understood when we pay
attention to their EoS as the pressure approaching zero. In the
QMDTD model, the energy density at zero pressure is finite, because
of the confinement mechanism, so the pressure vanishes much faster
than the strange hadronic matter. This property reduces the ability
of such strange quark star to support a dilute outer layer as in
strange hadronic star.

\begin{figure}
\resizebox{0.7\linewidth}{!}{\includegraphics*{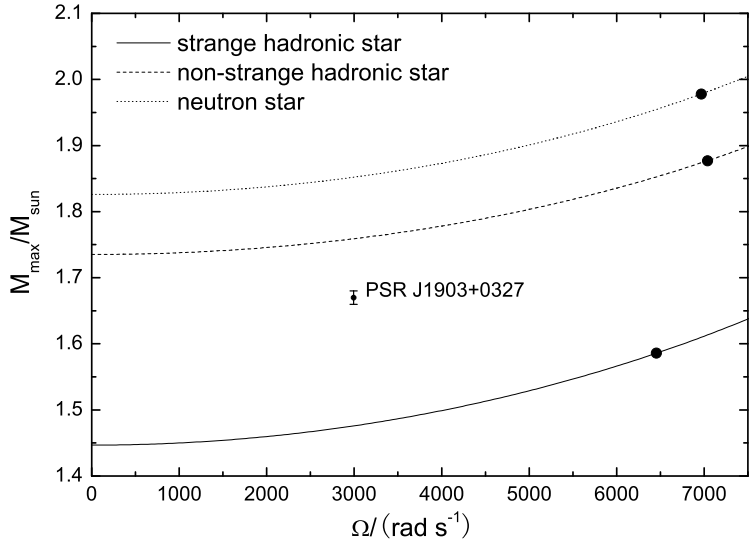}} \caption{The
curves of the maximum mass as functions of the angular velocity for
the three types of stars.} \label{MW}
\end{figure}

\begin{figure}
\resizebox{0.7\linewidth}{!}{\includegraphics*{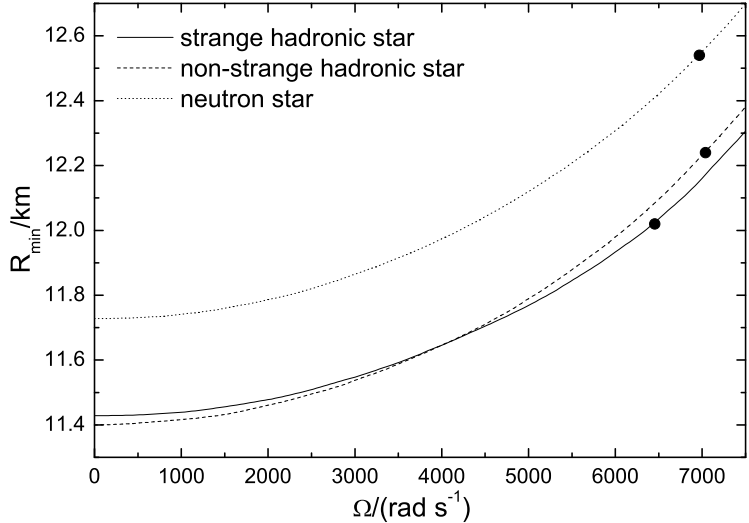}} \caption{The
curves of the minimum stable radius as functions of the angular
velocity for the three types of stars.} \label{RW}
\end{figure}

Different EoS can quantitatively change the mass and radius of a
star. The $M$-$R$ curves of these stars show that, at the same
condition, the strange quark star has the smallest radius, which is
not changed by about $1$km increase from the crust; while the
neutron star has the largest radius and mass. The non-strange
hadronic star has larger mass and radius than its strange relative.
The maximum mass of strange quark star is a little smaller than that
of the non-strange hadronic star. For example, at $\Omega=0$, the
neutron star has $M_{max}=1.8M_{sun}$, for the non-strange hadronic
star $M_{max}=1.7M_{sun}$, and the strange hadronic star has
$M_{max}=1.45M_{sun}$, while the strange quark star has
approximately $M_{max}=1.7M_{sun}$. These values are directly
related to the EoS of different matter. In general, this property
will not change: the steeper the $P$-$\rho$ curve is, the larger
mass the corresponding star will have. The rotation can not change
the order of the physical quantities of these types of stars,
either. To make this point clearer, we show the $M_{max}$ curves as
functions of $\Omega$ for different stars together in Fig.\ref{MW}.
Similarly, the $R_{min}$ curves are shown together in Fig.\ref{RW},
where we see that the minimum stellar radius of neutron star is
distinctly larger than the other two types.

\section{Summary and Discussions}

In summary, employing the EoS of the chiral SU(3) quark mean field
model and Hartle's method, we have studied the rotation effect on
different properties of the strange hadronic star, non-strange
hadronic star and neutron star. The most important effect of the
rotation is that it can increase the maximum mass, which is listed
clearly in Table \ref{mass}, where $\Omega_{max}$ is given by
Eq.(17) according to the maximum mass of static star. We see that
the range of the mass is significantly enlarged when the rotation is
considered. However, the maximum mass of the strange hadronic star
is still not large enough to explain the mass of PSR J1903+0327. As
can be seen from Fig.\ref{MW}, the point representing the mass and
the angular velocity of PSR J1903+0327 lies well below the $M_{max}$
curves of the non-strange hadronic star and the neutron star but far
above the curve of the strange hadronic star. So in the chiral SU(3)
quark mean field model this star can not be explained as a strange
hadronic star even the rotation effect is considered.

\begin{table}
\caption{The maximum mass of different stars with and without
rotation}
\begin{tabular}[t]{|c||c|c|c|}\hline
\centering
type of star & strange hadronic star & non-strange hadronic star & neutron star \\
\hline\hline
non-rotating $M_{max}$ & $1.45M_{sun}$ & $1.7M_{sun}$ & $1.8M_{sun}$ \\
\hline
$M_{max}$ at $\Omega_{max}$ & $1.586M_{sun}$ & $1.877M_{sun}$ & $1.978M_{sun}$ \\
\hline
$\Omega_{max}$ & $6454$rad/s & $7039$rad/s & $6968$rad/s \\
\hline
\end{tabular}\label{mass}
\end{table}

We can easily understand the low mass of the strange hadronic star
if we notice that its EoS is very soft. For other mean field
hadronic model with hyperons, for example, Refs.\cite{Hanauske:2000,
Schramm:2003}, the resulting maximum mass $M_{max}=1.52M_{sun}$ for
the strange hadronic star and $M_{max}=1.84M_{sun}$ for the neutron
star \cite{Hanauske:2000} at the static case are just a little
larger than those we presented here. Their results still confirm PSR
J1903+0327 is not a strange hadronic star, the same as our model.

If we want to increase the $M_{max}$ of strange hadronic stars, the
nonuniform rotation can be considered. As was pointed out in
Ref.\cite{Cook:dual}, the uniform rotation can just increase the
maximum stellar mass by about $20\%$ at most. Extend the discussion
from uniform rotation to differential rotation, the upper limit of
the stellar mass will increase remarkably \cite{Baumgarte:2000,
Lyford:2003,Morrison:2004}. Another possibility is to modify the
soft EoS of strange hadronic matter, for example, the mixing phase
of hadronic matter and quark matter. It has been proven that the EoS
can be significantly steepened and the maximum mass be elevated when
the mixing phase is considered \cite{Burgio:2003, Baldo:2003}. In
such case the deconfined quark becomes important and the interior
part of the star might be made of quark matter, that is the
so-called hybrid star \cite{Glendenning:1992b,Rosenhauer:1992,
Weber:1999}. It was shown that the hybrid star with a 2-flavor
color-superconducting quark core can support very heavy star
\cite{Alford:2007}. Another interesting question is about the
possibility of the ``quarkyonic matter" phase \cite{McLerran:2007,
Glozman:2008}. It is of interest to see the influence of such phase
on the $M$-$R$ relation if they can be attained in the core of the
compact star. However the discussion is beyond the chiral SU(3) mean
field model we considered in this paper, because this model cannot
provide the deconfinement mechanism.

\begin{acknowledgments}
We thank Dr. P. Wang for providing us the data of the EoS of the
chiral SU(3) quark mean field model. This work was supported in part
by NNSF of China. Shaoyu Yin is partially supported by the graduate
renovation foundation of Fudan university.
\end{acknowledgments}

\end{document}